\documentclass[nofootinbib,a4paper,aps,pra,showpacs,preprintnumbers]{revtex4-2}
\usepackage[T2A]{fontenc}
\usepackage{amssymb}
\usepackage{amsmath}
\usepackage{mathtools}
\usepackage{braket}
\usepackage{tikz}
\usepackage{graphicx}
\usepackage{epstopdf}
\usepackage[utf8]{inputenc}
\usepackage[russian, english]{babel}
\usepackage{url}
\usepackage{listings}
\usepackage{gensymb}
\usepackage{xcolor}
\usepackage{ulem}
\usepackage{enumitem}
\usepackage{comment}

\newlist{steps}{enumerate}{1}
\setlist[steps, 1]{label = Step \arabic*:}

\begin{document}
%
\title{QUANTUM ALGORITHM FOR REDUCING AMPLITUDES IN ORDER TO SEARCH AND FILTER DATA}
%

\author{Karina Zakharova$^1$*, Artem Chernikov$^1$, Sergey Sysoev$^2$}
%
%
%
\affiliation{${}^1$Saint Petersburg State University, Russia,\\
${}^2$Leonhard Euler International Mathematical Institute, Russia,\\
*reshetova.carina@yandex.ru}


\begin{abstract}
The method is introduced for fast data processing by reducing the probability amplitudes of undesirable elements. The algorithm has a mathematical description and circuit implementation on a quantum processor. The idea is to make a quick decision (down to a single iteration) based on the correspondence between the data and the desired result, with a probability proportionate to this correspondence. Our approach allows one to calibrate the circuit to control specified proportions.
\end{abstract}
\maketitle
\hyphenation{}
\section{Introduction}
\begin{figure}
\centering
\includegraphics[width=1\textwidth]{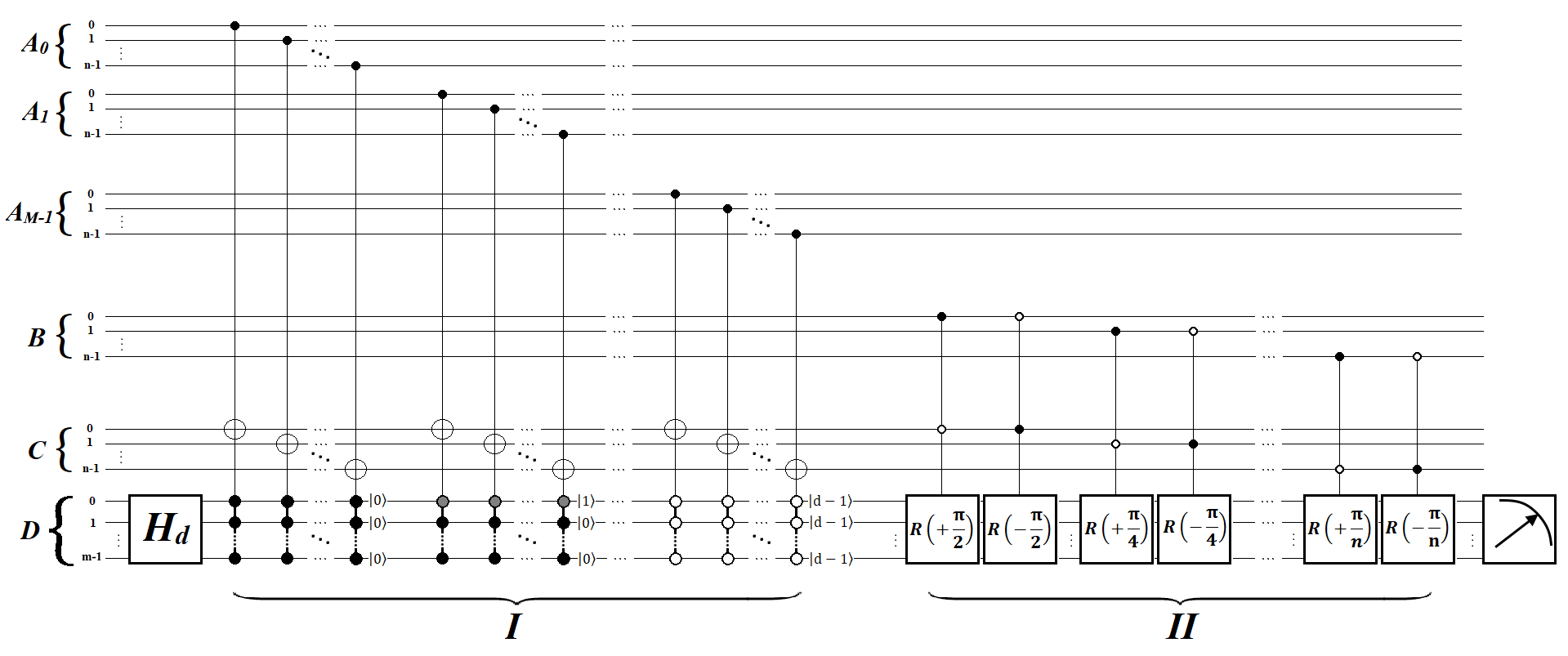}
\caption{Previously proposed algorithm scheme}\label{fig:1PropSch}
\end{figure}

\begin{figure}
\centering
\includegraphics[width=0.9\linewidth]{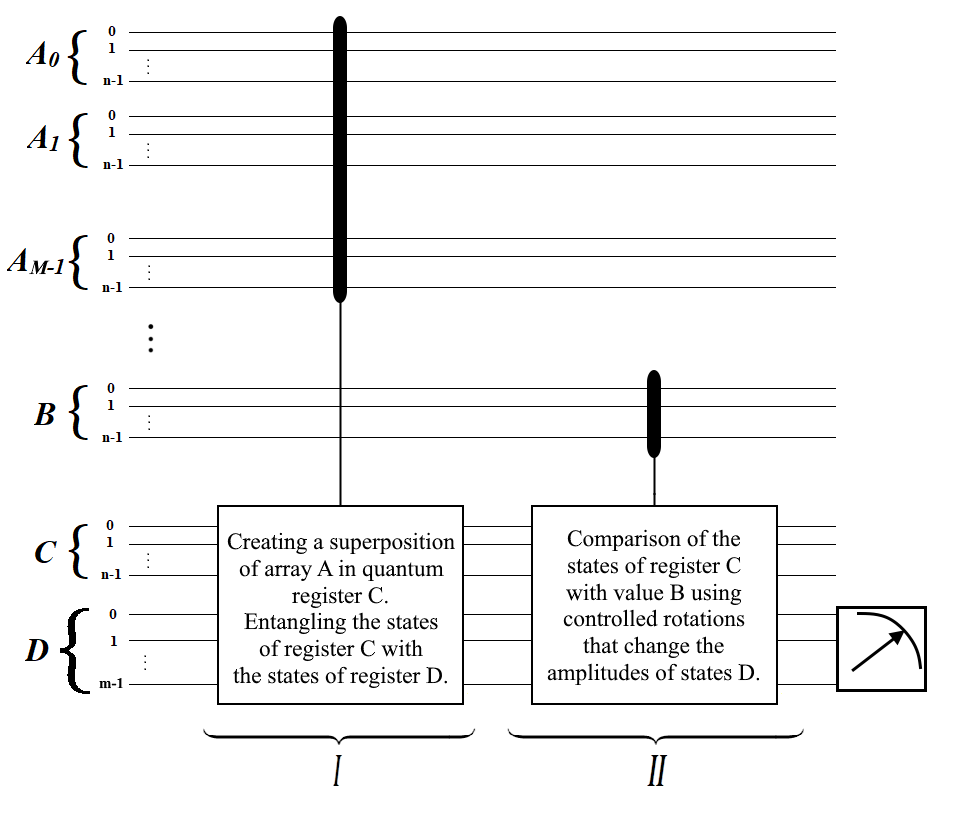}
\caption{Generalized scheme of the method}\label{fig:2GenSch}
\end{figure}

In \cite{8QARN}, a method is proposed to find the closest values. The scheme and algorithm of this method are shown in Figure \ref{fig:1PropSch}. The idea is to find values closest to a specified $B$, of the same $n$-bit length, in an array $A$ containing $M$ elements, each with $n$ bits, about which there is no information available (Figure \ref {fig:2GenSch}).
In the first stage of the process, a local copy of the original data (array $A$) is stored in an $n$-qubit quantum register $C$. Each element $C_i$ of the register is entangled with a counter, which introduces it into a superposition of all possible quantum states using Hadamard gates. The counter can be composed of qubits or qudits (\cite{4Amit}, \cite{5Wang}, \cite{6Mogos}, \cite{7Carretta}); the main requirement is that it has $M$ states in the superposition, with $M = d^m$, where $d$ is the number of levels for one qubit and $m$ is the number of qudits. Therefore, array $A$ appears in a superposition within register $C$, allowing for processing of all its elements simultaneously.
Next, the search for the closest to $B$ is carried out using a bitwise comparison with each possible state $C_i$. The difference between $C_i$ and $B$ can be expressed as a change in the probability amplitude of $D_i$, which is entangled with $C_i$. A rotation matrix, proposed in \cite{8QARN}, is also presented below:
\begin{equation}
\left(\begin{array}{cccc}
\cos \frac{\varphi}{2} & \pm \frac{\sin \frac{\varphi}{2}}{\sqrt{2^m-1}} & \cdots & \pm \frac{\sin \frac{\varphi}{2}}{\sqrt{2^m-1}} \\
\pm \frac{\sin \frac{\varphi}{2}}{\sqrt{2^m-1}} & \cos \frac{\varphi}{2} & \cdots & \pm \frac{\sin \frac{\varphi}{2}}{\sqrt{2^m-1}} \\
\vdots & \vdots & \ddots & \vdots \\
\pm \frac{\sin \frac{\varphi}{2}}{\sqrt{2^m-1}} & \pm \frac{\sin \frac{\varphi}{2}}{\sqrt{2^m-1}} & \cdots & \cos \frac{\varphi}{2}
\end{array}\right)
\end{equation}
The amplitude of each $|D_i\rangle$ state is altered by controlled rotations in space, depending on the difference between $C_i - B$. The key difference from Grover's approach is that there are no labeled elements in the array (recall that we have no knowledge of the input data, including whether or not there is an exact match between $A_k$ and $B$, the number of values that are close to $B$, or the distance between them), and we aim to reduce the probability of undesirable outcomes rather than increasing the probability of desirable ones. By default, we propose encoding the difference between the highest qubits as $\frac{\pi}{2}$, the next as $\frac{\pi}{4}$, etc. Recall that since the entire array is in a superposition state, comparisons between $B$ and all possible states of $C$ occur simultaneously.
The greater the difference between $C_i$ and $B$, the smaller the $\cos\frac{\varphi}{2}$ for state $D_i$. The entire "subtracted" probability is then evenly distributed among other possible states of the counter $D$ ($\frac{\sin \frac{\varphi}{2}}{\sqrt{2^m-1}}$). It is essential to ensure that the sum of rotation angles does not exceed $\pi$, otherwise there is a risk of overshooting and causing the amplitude to begin to twist in opposite direction, similar to the case of unnecessary iterations in Grover's algorithm \cite{1Grov}.
So, the main differences from the latter are the absence of an ancilla qubit \cite{2Bras,3Prak} with a "recording" of the probability amplitude directly into the counter $D$ (and not in the form of phases), a weakening of the probabilities of "bad" values (rather than an amplification of "good" ones), the absence of the need for an exact match $C_i = B$ as a criterion for successful search and the main difference is the absence of a black box, or oracle. The proposed scheme is transparent and can be implemented on a quantum processor. At the Supercomputing Days 2024 conference, as part of a poster presentation, our scientific subgroup announced the method of implementing the proposed algorithm, as well as the results of the first studies.
In this article, we offer a detailed report on the work done as part of a comprehensive study on the developed method. We describe its capabilities, disadvantages and goals for future application. It's also worth noting that our initial goal wasn't to maximize the likelihood of finding only the closest value. Instead, we aimed to be able to efficiently process the data and, with high certainty, obtain an optimal result. From the outset, the ability to run the algorithm multiple times and get several (potentially different or identical) results was seen as an advantage, allowing for a deeper analysis of the results based on other criteria later on.
\section{Implementation of the scheme. The possibilities and limitations of a single call}
\begin{figure}
\centering
\includegraphics[width=0.8\linewidth]{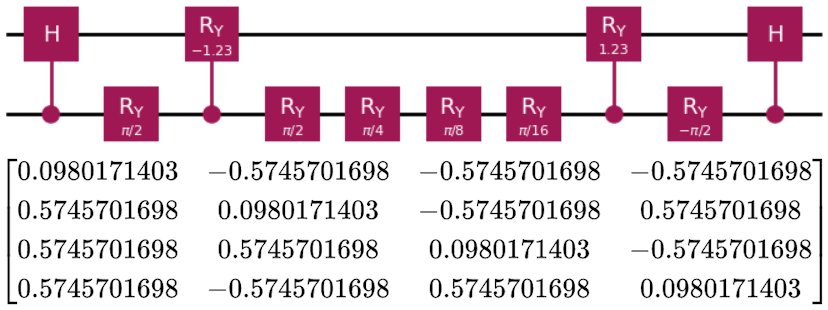}
\caption{Example of an operator for rotating a 4-dimensional state by an angle equal to $\frac{15\pi}{16}$, which can encode a difference equal to the value of 15 (or 1111 in binary)}\label{fig:3_1111}
\end{figure}

\begin{figure}
\centering
\includegraphics[width=0.9\linewidth]{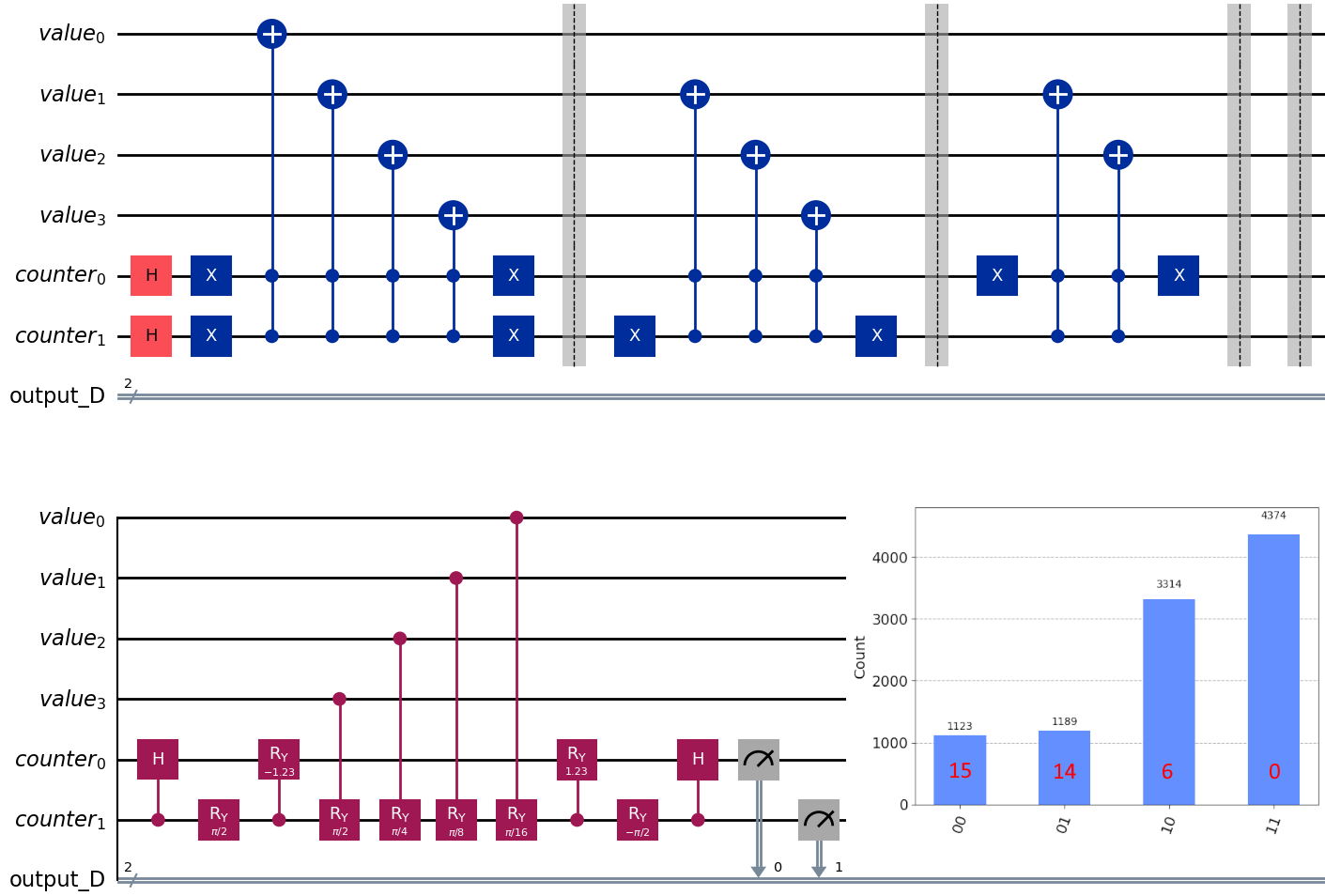}
\caption{Example of a scheme for finding the closest to zero in an array [15 14 6 0]}\label{fig:4_1mes}
\end{figure}

\begin{figure}
\centering
\includegraphics[width=0.8\linewidth]{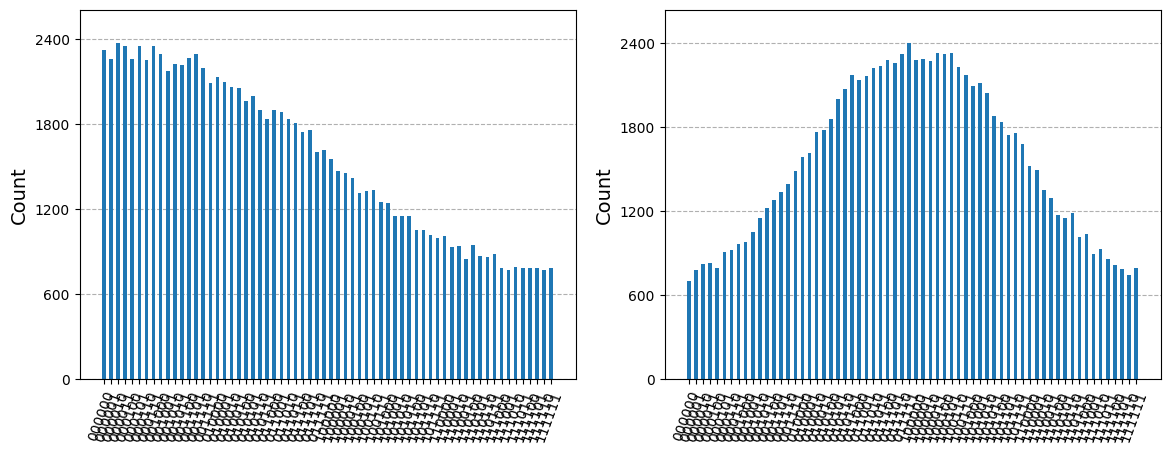}
\caption{The results for 100,000 measurements of a single call to the algorithm for arrays of 64 elements [0 1 2 3... 61 62 63] and [63 61 59... 3 1 0 2 4... 60 62], respectively}\label{fig:5_1mes}
\end{figure}

As mentioned previously, a general framework for implementing the algorithm and a matrix representation of rotations were proposed in reference \cite{8QARN}. In this paper, we have successfully derived the unitary form of this operator and implementated it using quantum gates.
Figure \ref{fig:3_1111} illustrates an example of a rotation of a two-qubit state, with a numerical representation of the operator. Figure \ref{fig:4_1mes} shows an implementation circuit for these operators.
For the purpose of simplicity in research and to simplify the scheme, all elements of array $C$ will be compared to zero ($B=0$).
In the upper part of the circuit, an array $C = [15, 14, 6, 0]$ with a size of 4 qubits is entangled with a counter $D$ with a size of 2 qubits.
The null element $C_0=|1111\rangle$ (corresponding to a value of 15) is entangled with the state $D_0 = |00\rangle$. The first element $C_1 = |1110\rangle$ (14) is entangled with $D_1 = |01\rangle$. The second $C_2 = |0110\rangle$ (6) is entangled with $D_2 = |10\rangle$, and the third $C_3 = |0000\rangle$ (0) is entangled with $D_3 = |11\rangle$. The amplitudes of each state are the same and equal $1/2$.
In the lower half of the circuit, a series of controlled rotations are applied to the qubits. If the highest qubit of a given element of $C_k$ is 1, then the state $D_k$ is rotated by $\frac{\pi}{2}$, then $\frac{\pi}{4}$, and so on, depending on the value of the qubit. Therefore, the larger the value of an element of $C_k$, the greater the angle by which the corresponding state $D_k$ will be rotated.
Obviously, since $C_0=15$ is the farthest from zero, the matrix corresponding to the initial amplitude of $D_0$ (equal to 1/2) will "remove" the component corresponding to the maximum rotation ($\cos{\frac{\frac{\pi}{2}+\frac{\pi}{4}+\frac{\pi}{8}+\frac{\pi}{16}}{2}}$) and "distribute" it equally among the states $D_1$, $D_2$, and $D_3$ ($\frac{\sin{\frac{\frac{15\pi}{16}}{2}}}{\sqrt{4-1}}$).
Because all four elements are in superposition within register $C$ (as well as all counter values in register $D$, respectively), in a series of rotations, each element of $C$ is simultaneously compared to zero. Therefore, state $D_0$ also "acquires" angles $\frac{14\pi}{16}$, $\frac{6\pi}{16}$, and $\frac{0\pi}{16}$ from $D_1$, $D_2$, and $D_3$, respectively.
Each element seems to move a portion of amplitude away from its own state towards other states. The larger the element, the smaller portion it retains for itself, while also taking something from others. Thus, the final redistribution depends on the specific values of all data.
At the end of the process, counter $D$ is measured and its value indicates element number $C_j$.
The graph in the lower right corner displays the results of 10,000 measurements.
In this instance, the algorithm involves loading an array and performing rotations, which corresponds to a computational complexity of $O(1)$ in the Grover sense. The simulation equations for this scheme are as follows in two stages:
\begin{equation}
\frac{1}{2}(|C_0\rangle+|C_1\rangle|01\rangle+|C_2\rangle|10\rangle+|C_3\rangle|11\rangle)\nonumber\
\end{equation}\
\begin{equation}
\overset{Rotation}{\rightarrow}\nonumber\\
\end{equation}
\begin{equation}
\frac{1}{2}[|C_0\rangle(a_0|00\rangle+b_0(|01\rangle+|10\rangle+|11\rangle))+\nonumber\\
\end{equation}
\begin{equation}
+|C_1\rangle(a_1|01\rangle+b_1(-(00\rangle+|10\rangle+|11\rangle)+\nonumber\\
\end{equation}
\begin{equation}
+|C_2\rangle(a_2|10\rangle+b_2(-|00\rangle-|01\rangle+|11\rangle)+\nonumber\\
\end{equation}
\begin{equation}
+|C_3\rangle(a_3|11\rangle+b_3(-|00\rangle+|01\rangle-|10\rangle)]
\end{equation}
The values $a_i$ and $b_i$ represent the total rotation coefficients, which depend on the element $C_i$, as illustrated in Figures \ref{fig:3_1111} and \ref{fig:4_1mes}. The matrix can be represented in the following general form:
\begin{equation}
A=\left(\begin{array}{cccc}
a & -b & -b & -b \\
b & a & -b & b \\
b & b & a & -b \\
b & -b & b & a
\end{array}\right), a = \cos \frac{\varphi}{2}, b = \frac{\sin \frac{\varphi}{2}}{\sqrt{2^2-1}}
\end{equation}
Figure \ref{fig:5_1mes} presents measurement graphs for a circuit with larger arrays consisting of 64 elements ranging from 0 to 63 and their various positions within the array. The amplitude values on the graphs confirm a consistent distribution of probabilities for each execution of the algorithm. Previous studies have shown that the maximum possible increase in the probability of a particular element, regardless of the number of elements $M$, is a factor of 2.
\section{Matrix constraints and using the decoherence}
\begin{figure}
\centering
\includegraphics[width=0.7\linewidth]{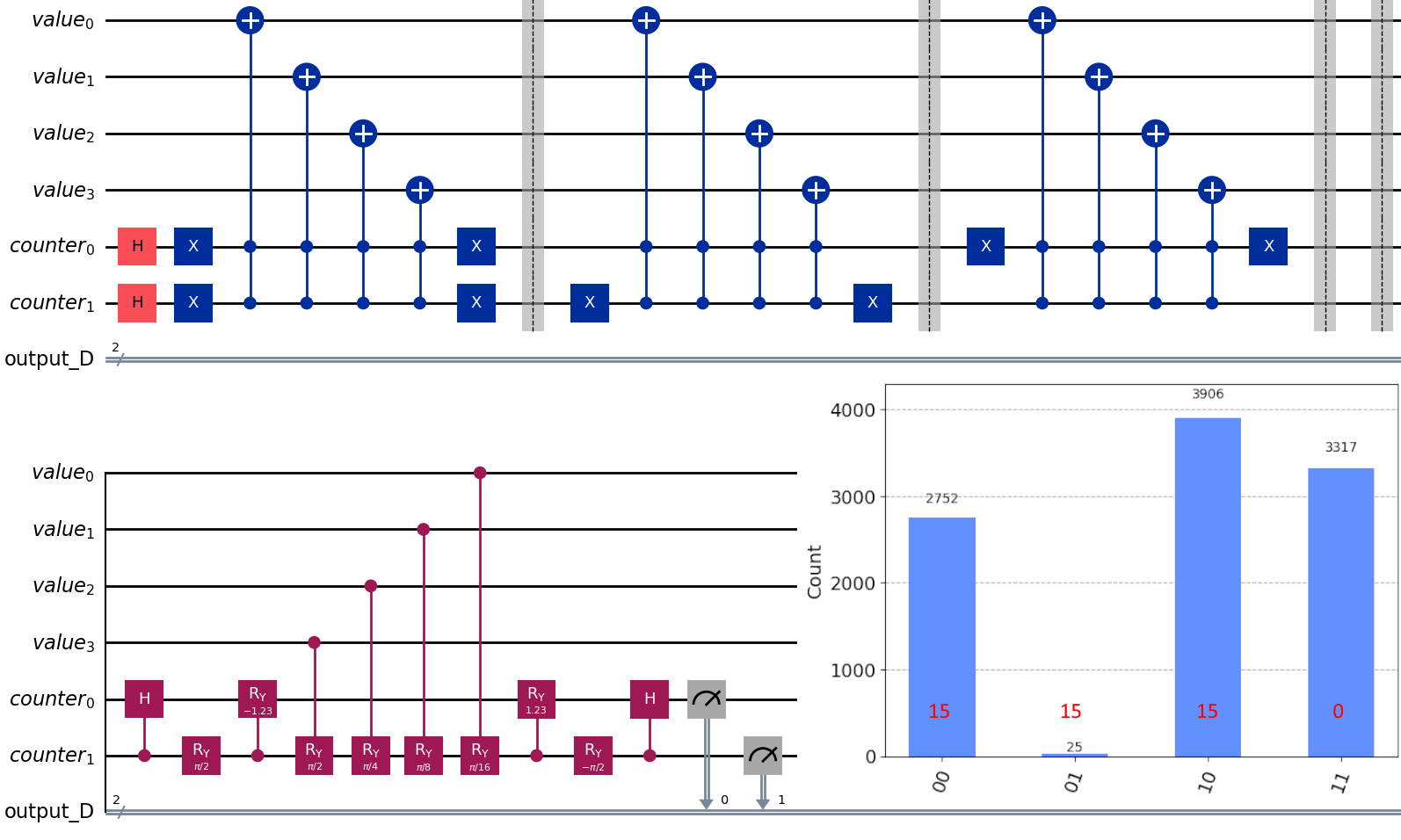}
\caption{Example of a scheme for finding the closest to zero in an array [15 15 15 0]}\label{fig:6_repeat}
\end{figure}

\begin{figure}
\centering
\includegraphics[width=0.8\linewidth]{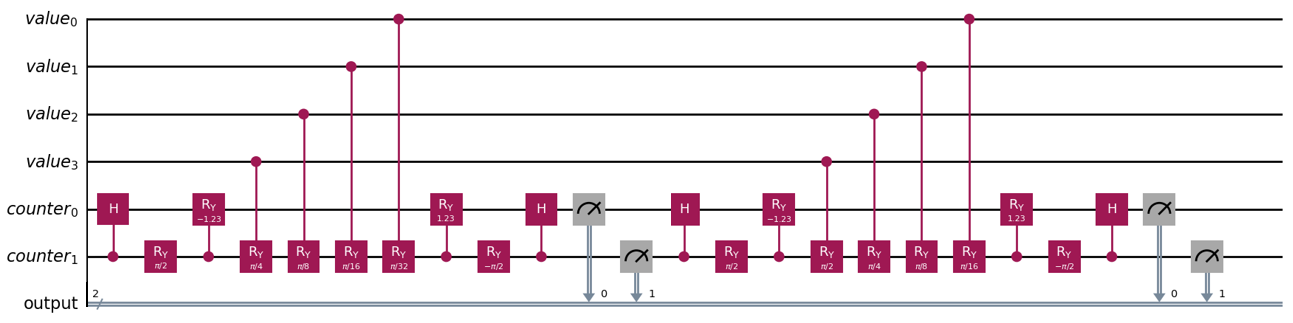}
\caption{Duplicated rotations and measurements without reloading the array}\label{fig:7_re-meas}
\end{figure}

\begin{figure}
\centering
\includegraphics[width=0.9\linewidth]{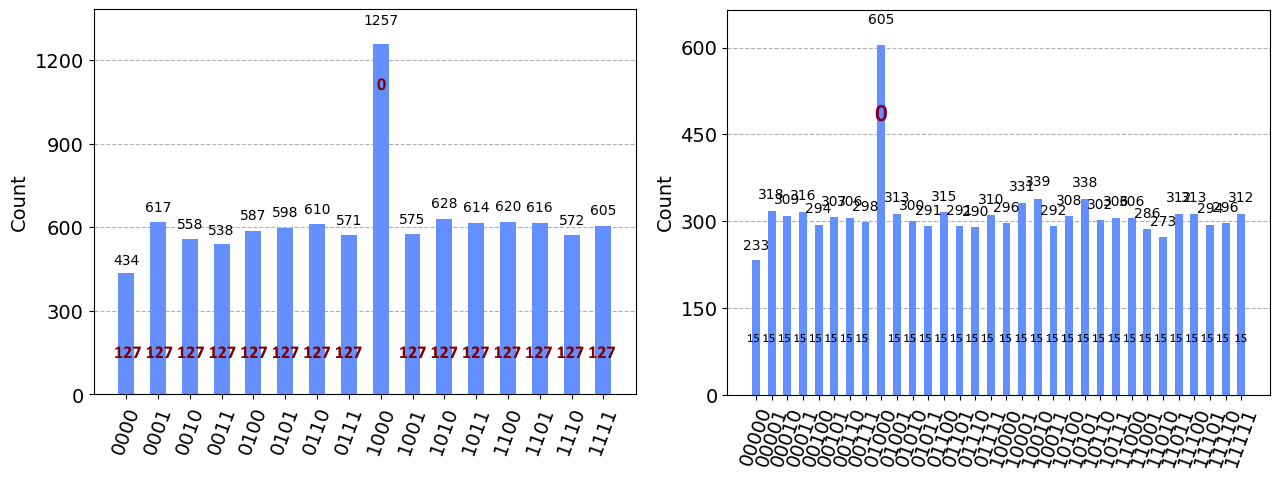}
\caption{Re-measurement results for arrays of fifteen and thirty-one identical elements (equal to 127 and 15, respectively) and an eighth element equal to zero}\label{fig:8_2meas}
\end{figure}

Previously, we considered arrays with no duplicate elements. However, if, for example, $C_0 = C_2$, the counters associated with them can be added together ($|C_0\rangle(a_0|00\rangle+b_0(|01\rangle+|10\rangle+|11\rangle))+|C_2\rangle(a_2|10\rangle+b_2(-|00\rangle-|01\rangle+|11\rangle)$).
And since, depending on the position in the matrix, the coefficients $a$ and $b$ have different signs, the results differ from those required, as demonstrated in the example in Figure \ref{fig:6_repeat}.
Decoherence proved to be an unanticipated solution, specifically, a measurement prior to repeated rotation and a subsequent measurement, as illustrated in Figure \ref{fig:7_re-meas}.
This can be explained by the fact that after the first sequence of rotations, each state of counter $D$ is entangled with all other states of $C$. After the first measurement, the repeated rotation operator is applied to a specific state of $D_i$. Therefore, for 10,000 measurements (see Figure \ref{fig:8_2meas}), the statistics are successfully recovered.
At the same time, a maximum twofold increase in amplitude is also achieved, irrespective of the amount of data and its values. This method is rather heuristic in nature, but it is worth mentioning for future research into possible use of decoherence.
\section{Using the null element}
\begin{figure}
\centering
\includegraphics[width=0.9\linewidth]{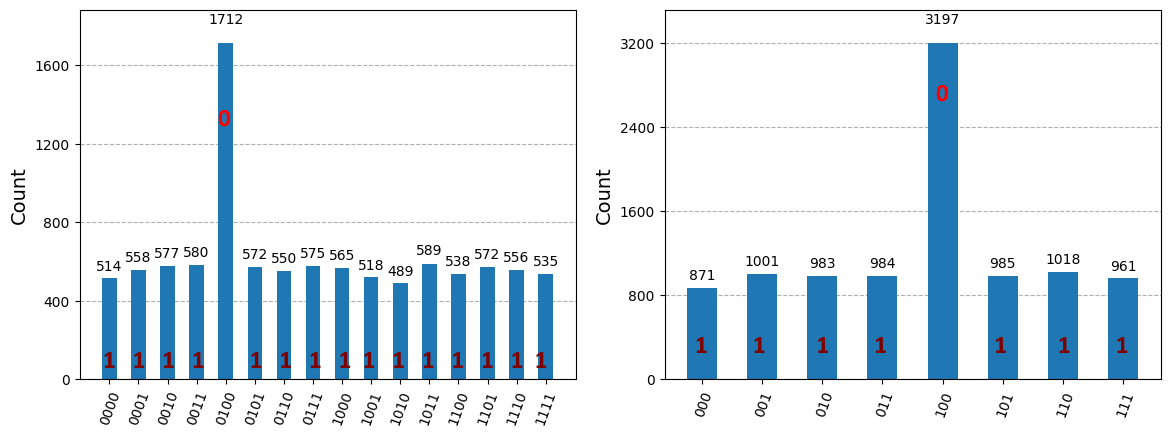}
\caption{Re-rotation results for arrays of sixteen and eight elements, respectively, with a single zero in each}\label{fig:10_null}
\end{figure}

As can be observed from the matrix representation of the rotation operator, the sign constancy only holds for the null element, which explains the increase in its amplitude (as previously mentioned, with many repeated elements, “entangled” amplitudes can now accumulate and cancel each other due to differences in signs). We propose an alternative solution to this issue, utilizing the effect of “dragging” the amplitude by a null element.
In order to accomplish this, we adjust the matrix that executes a single rotation by $\pi$ on all elements of $C_i$ that do not equal the desired $B$.
This is a significant point, as we now assume that there is a perfect match in the original array and reduce all other values by the maximum angle, irrespective of their proximity to $B$, by the angle $\pi$. Below is an illustration of a rotation matrix used to locate an exact match:
\begin{equation}
\left(\begin{array}{cccccccc}
0 & -\frac{\sqrt{7}}{7} & -\frac{\sqrt{7}}{7} & -\frac{\sqrt{7}}{7} & -\frac{\sqrt{7}}{7} &-\frac{\sqrt{7}}{7} & -\frac{\sqrt{7}}{7} & -\frac{\sqrt{7}}{7}\\
\frac{\sqrt{7}}{7} & 0 & -\frac{\sqrt{7}}{7} & \frac{\sqrt{7}}{7} & -\frac{\sqrt{7}}{7} & \frac{\sqrt{7}}{7} & -\frac{\sqrt{7}}{7} & \frac{\sqrt{7}}{7} \\
\frac{\sqrt{7}}{7} & \frac{\sqrt{7}}{7} & 0 & -\frac{\sqrt{7}}{7} & -\frac{\sqrt{7}}{7} & -\frac{\sqrt{7}}{7} & \frac{\sqrt{7}}{7} & \frac{\sqrt{7}}{7} \\
\frac{\sqrt{7}}{7} & -\frac{\sqrt{7}}{7} &\frac{\sqrt{7}}{7} & 0 & -\frac{\sqrt{7}}{7} & \frac{\sqrt{7}}{7} & \frac{\sqrt{7}}{7} & -\frac{\sqrt{7}}{7} \\
\frac{\sqrt{7}}{7} & \frac{\sqrt{7}}{7} & \frac{\sqrt{7}}{7} & \frac{\sqrt{7}}{7} & 0 &  -\frac{\sqrt{7}}{7} & -\frac{\sqrt{7}}{7} & -\frac{\sqrt{7}}{7}\\
\frac{\sqrt{7}}{7} & -\frac{\sqrt{7}}{7} & \frac{\sqrt{7}}{7} & -\frac{\sqrt{7}}{7} & \frac{\sqrt{7}}{7} & 0 & -\frac{\sqrt{7}}{7} & \frac{\sqrt{7}}{7} \\
\frac{\sqrt{7}}{7} & \frac{\sqrt{7}}{7} & -\frac{\sqrt{7}}{7} & -\frac{\sqrt{7}}{7} & \frac{\sqrt{7}}{7} & \frac{\sqrt{7}}{7}  & 0 & -\frac{\sqrt{7}}{7} \\
\frac{\sqrt{7}}{7} & -\frac{\sqrt{7}}{7} & -\frac{\sqrt{7}}{7} & \frac{\sqrt{7}}{7} & \frac{\sqrt{7}}{7} & -\frac{\sqrt{7}}{7} & \frac{\sqrt{7}}{7} & 0 \\
\end{array}\right)
\end{equation}
To further simplify the process, we will search for a single zero among many units. After loading an array (entangling register $C$ with counter $D$) and applying a rotation operator, it will be necessary to mark the null element (state $|D_0\rangle$). For this purpose,additional ancillary qubits will be required.
\begin{figure}
\centering
\includegraphics[width=1\linewidth]{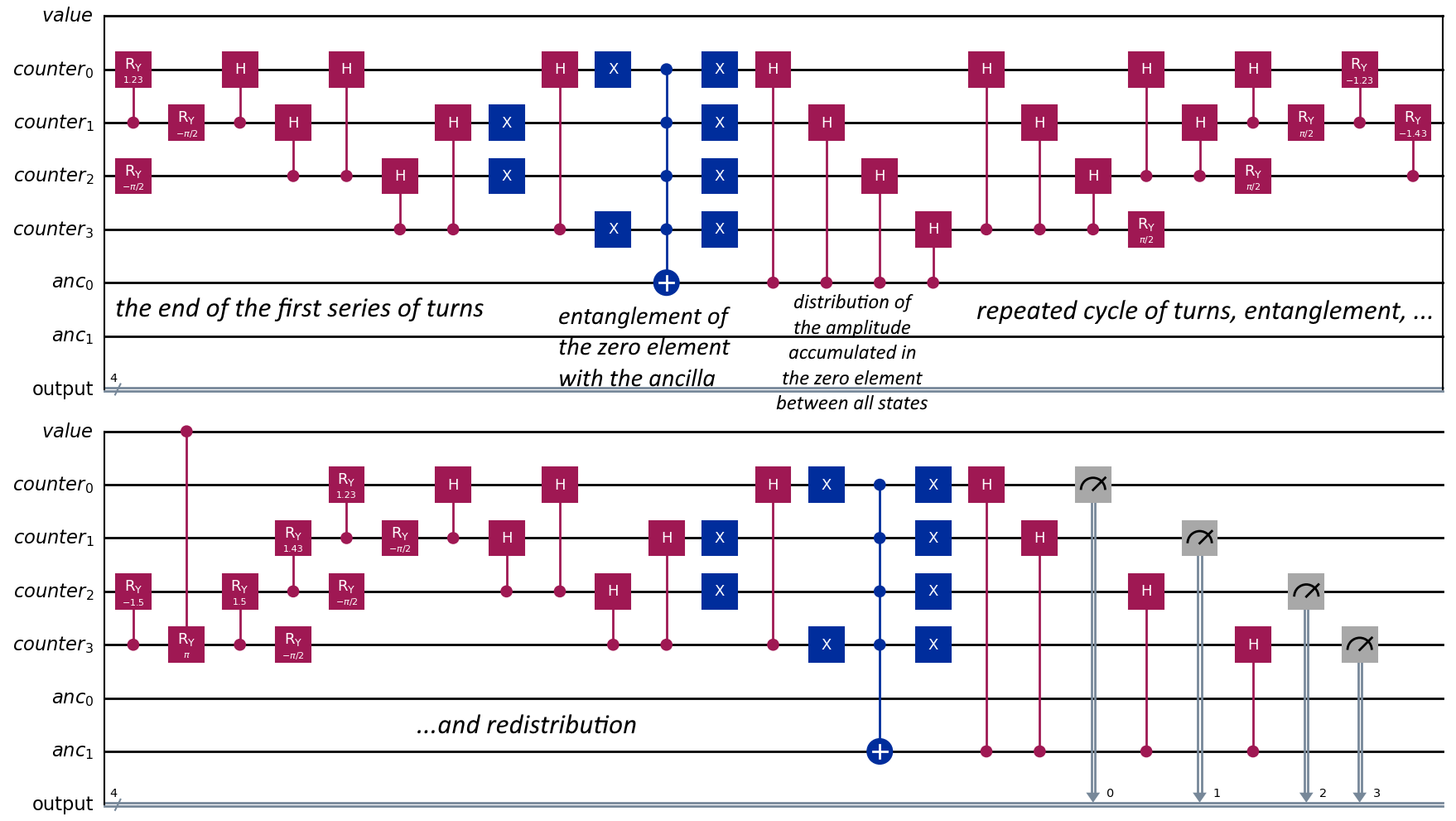}
\caption{Re-rotation scheme using the null item}\label{fig:9_rerot}
\end{figure}
In Figure \ref{fig:9_rerot}, this step is implemented using NOT gates and multiple CNOT gates, which are colored blue. Next, using controlled Hadamard gates, any erroneously accumulated amplitude in the zero state is distributed among all states of the register, after which a further cycle of rotations, measurements, and redistributions is planned before measuring the register $D$.
Therefore, the desired element does not provide its amplitude at each iteration, but only receives it. At this time, the maximum amplitude increase is not 2-fold, but 3-fold compared to the initial weighted value, as was previously the case, regardless of the amount of data (Figure \ref{fig:10_null}).
\subsection{Iterative use of the null element}
\begin{figure}
\centering
\includegraphics[width=0.8\linewidth]{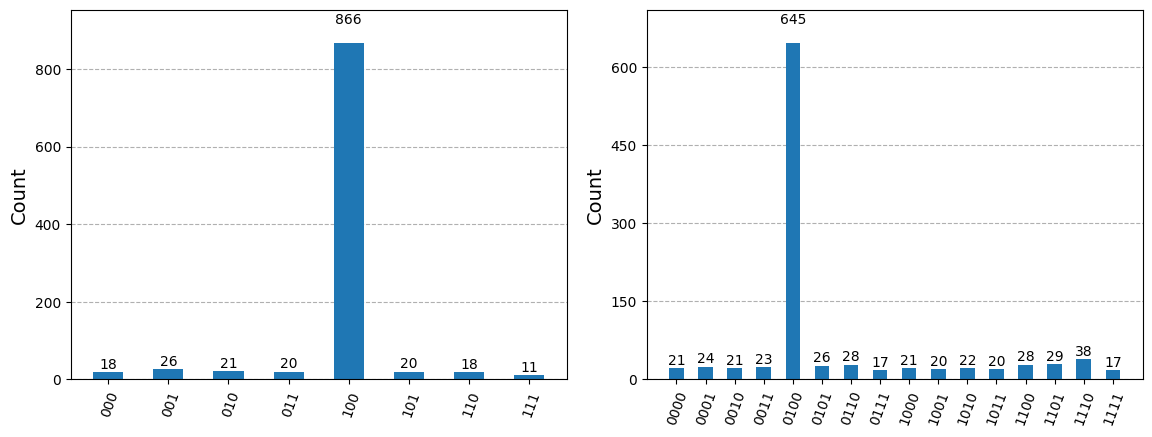}
\caption{Eight calls the algorithm for 2 arrays of 8 and 16 elements, respectively}\label{fig:11_8loads}
\end{figure}

\begin{figure}
\centering
\includegraphics[width=1\linewidth]{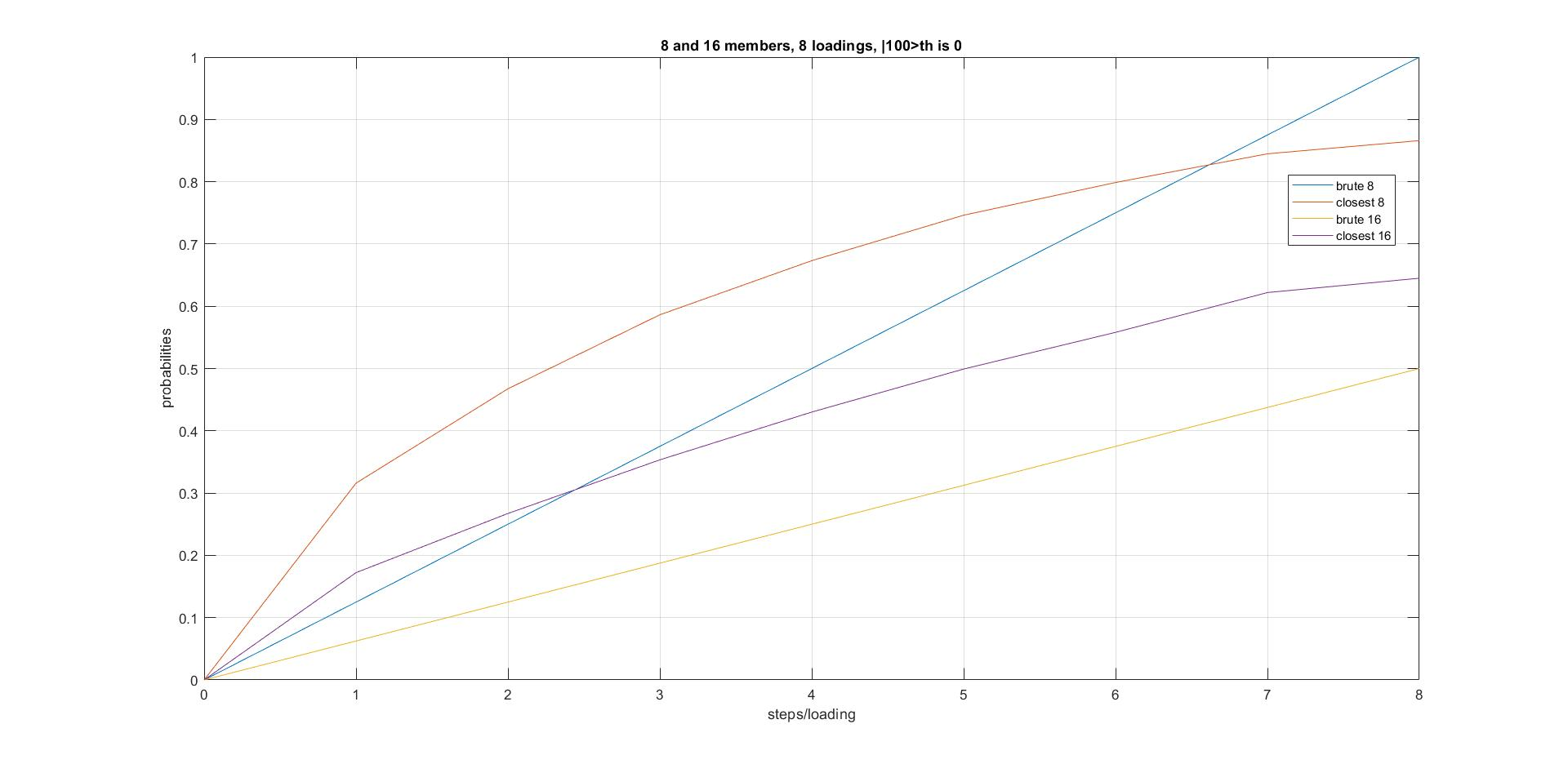}
\caption{The probability of finding the desired element increases depending on the number of iterations}\label{fig:12_brut}
\end{figure}

The method of using a null element as a guaranteed buffer to concentrate amplitudes allowes us to develop the idea of increasing amplitudes with each iteration of the algorithm.
This approach requires not only an additional ancillary qubit at each iteration, as we mentioned previously when discussing searching for repeated elements using the "parasitic" amplitude provided by the null element, but also an extra register $C'$ for data reloading.
The process of reloading data is carried out by re-entangling the counter states with each element in $C'$, while the counter states no longer have a uniform superposition of amplitudes, but probabilities that have been redistributed from the previous iteration.
Figure \ref{fig:11_8loads} illustrates the results of the build-up process for 8 calls. On the left, we see the familiar array of 8 elements (see Figure \ref{fig:10_null}), and on the right, a new array of 16 elements has been generated. In each of these arrays, we are searching for the fourth element that equals zero.
This experiment demonstrated that iteration is feasible, but the problem of uncomputing the ancilla (to conserve resources) remains, and the increase in amplitude is not intense enough. The graph (Figure \ref{fig:12_brut}) compares the change in the probability of detecting the target element at each iteration for our approach and the brute force method. The red and blue lines represent an array of 8 elements, and the purple and yellow lines represent an array of 16 elements.
\section{Iterativity in general}
\begin{figure}
\centering
\includegraphics[width=1\linewidth]{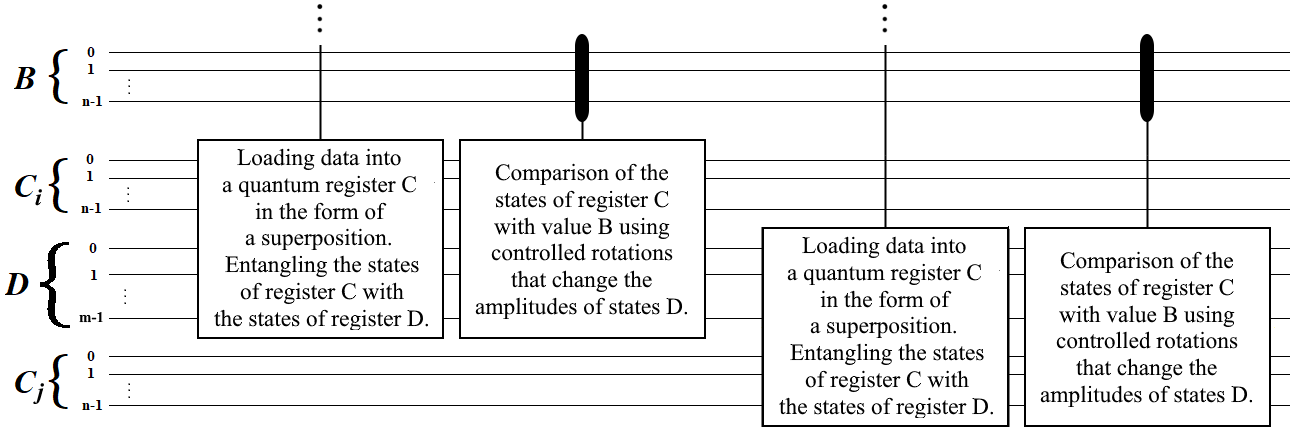}
\caption{The general scheme of the example of two subsequent iterations (calling with reloading an array) of the algorithm}\label{fig:13_iter}
\end{figure}

\begin{figure}
\centering
\includegraphics[width=0.9\linewidth]{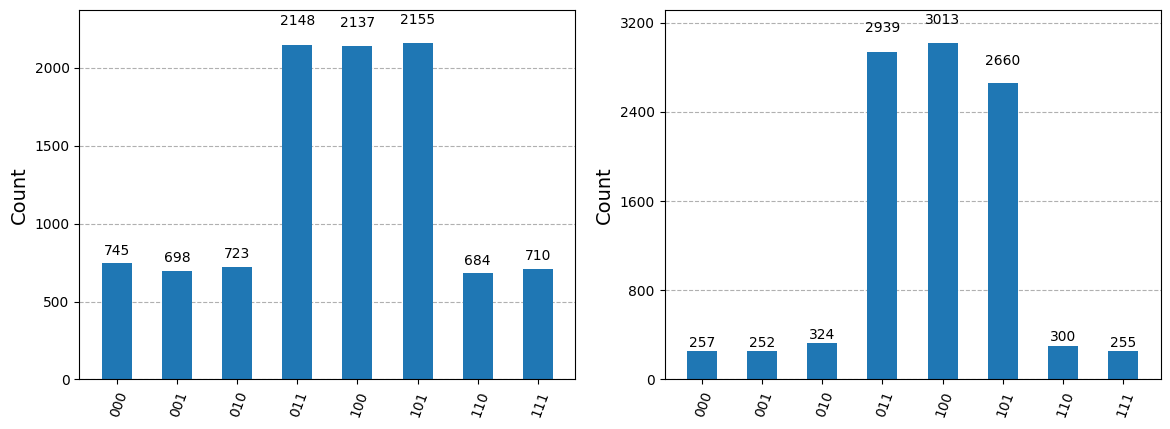}
\caption{The general scheme of the example of two subsequent iterations (calling with reloading an array) of the algorithm}\label{fig:14_gen_iter}
\end{figure}

In the general case, iteration is also possible when searching for the nearest element without focusing on a specific match. If we assume that the source data does not contain repeated elements, then we can avoid the use of an additional ancillary register and, consequently, the need for markinf the zero element, controlled Hadamard gates, and additional chains of rotations, as shown in Figure \ref{fig:9_rerot}. 
Instead, there will only be additional data loading for each iteration, which will involve entanglement with a counter storing the accumulated distribution of amplitudes from previous iterations, as illustrated in Figure \ref{fig:13_iter}.
At the same time, during each iteration, it is possible to adjust the rotation angles and control qubits that activate them. Figure \ref{fig:14_gen_iter} illustrates the results of an experiment with an array of eight six-qubit elements after the first iteration (left graph) and after three iterations (right graph).
The left graph displays the results after applying a single rotation by the highest qubit with an angle of $\pi$, which maximizes the elimination of all elements greater than 31 in value, which in fact turned out to be a process of amplitude sorting of the elements into those greater or lesser than the average value of 31.
In the graph on the right, this effect is maximized through experimentally determined iterations: the initial iteration performs the series of rotations described earlier across all qubits starting from $\frac{\pi}{2}$ from the highest qubit (as shown in Figure \ref{fig:4_1mes}) followed by two additional iterations, as shown in the graph on the left, where only the highest qubit in register $C$ rotates $D$ by $\pi$.
This process can be likened to applying an equalizer to an audio signal, where no prior knowledge of the incoming data is available, but one can attempt to influence the signal as much as possible by adjusting certain parameters in order to maximize the desired output. This involves "squeezing" out specific frequency bands in order to achieve the desired effect.
\section{Anti-search. Filter mode}
\begin{figure}
\centering
\includegraphics[width=1\linewidth]{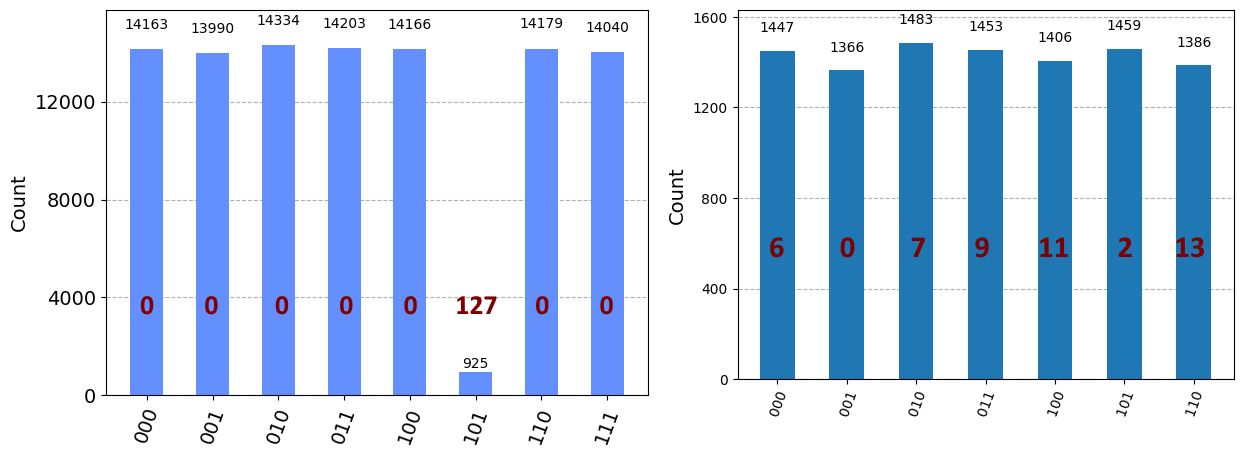}
\caption{Filtering the value on the left as far as possible from the desired (zero) value and excluding an exact match with the value of 15 on the right}\label{fig:15_Filt}
\end{figure}

Until now, we have discussed a controlled amplitude reduction technique to enhance desired values. In the final section, we propose using the suggested algorithm for its intended use, namely, attenuating until complete removal of undesired data.
In the left graph (see Figure \ref{fig:15_Filt}), we applied the method of finding the nearest value with repeated measurements in the case where there was only one "outlier" in the dataset. In this scenario, the amplitude of the undesired signal decreased significantly.
In the right graph (Figure \ref{fig:15_Filt}), we examine the array $[6, 0, 7, 9, 11, 2, 13, 15]$, and the goal is to remove the last element ($C_7 = 15$).
As can be seen, the number $D_7 = 111$ is not present in the sequence, which implies that the probability of its occurrence has been reduced to zero.
\section{Conclusions}
The tool allows users to control the suppression of unwanted values by sorting them by amplitude, with the "worst" elements having the smallest amplitudes. The rotation angles of the data array, on which the degree of suppression depends, are adjusted within specified limits. The amplitude of the nearest element can be increased up to a maximum of 2 or 3 times in a single call to the algorithm ($O(1)$), depending on the search criteria and the presence of duplicate data). It is possible to gradually increase an amplitude at each iteration with varying success compared to brute force method.
With a large amount of data, the described effects may be insignificant. However, in situations where it is necessary to quickly get the optimal result (after a single call) or $k$ optimal results after $k$ calls, the proposed algorithm can be used for quick decision-making. In addition, if you often need to process a small amount of data, this method allows you to avoid multiple comparisons with a fairly good result (for example, using a median filter).
And especially if there is no prior knowledge about the incoming signal, but it is necessary to exert as much influence as possible in a specific direction, the algorithm can be configured as an equalizer. Furthermore, if it is required to eliminate any components, the proposed method has a high probability of filtering them out in a single iteration.
All the experiments described can be accessed via the following link:
\url{https://colab.research.google.com/drive/18p3hsSwZRs7M2_0j7QuFmEFGadxY7ObY?usp=sharing}
%
%

\end{document}